\documentclass[11pt]{article}
\usepackage{amsmath, amsfonts, amssymb}
\usepackage{graphics}

\makeatletter
\renewcommand{\@oddhead}{\it \footnotesize Astromony Letters,
 Vol.31, 2005, pp.80-87 \hfil}
 \makeatother
\begin{document}
\newpage
\begin{center}
\huge \rm  Leading Wave as a Component  of the Spiral Structure
of  the Galaxy
\par \vspace{30pt} \small \it \copyright 2005 Pleiades Publishing
Inc.
\par \vspace{30pt} \large \rm A.M.Mel'nik
\par \it anna@sai.msu.ru
\par  \small \it Sternberg Astronomical Institute,
Moscow, Russia
\end{center}
\renewcommand{\abstractname}{}
\begin {abstract}
The spiral pattern of the Galaxy identified by analyzing the
kinematics of young stars within  3~kpc  of the Sun is Fourier
decomposed into spiral harmonics. The spiral pattern of the
Galaxy is shown to be representable as a superposition of
trailing and  leading waves with  interarm distances of
$\lambda=1.8\pm0.4$~kpc and $\lambda=4\pm2$~kpc, respectively.
Shock waves are probably present only in the portions of the
trailing spiral pattern where it crosses the crest of the leading
wave. The small interarm distance of the trailing spiral wave
($\lambda=1.8$~kpc) can be explained by its evolution --- by the
decrease in the interarm distance as the wave is displaced toward
the inner Lindblad resonance. The Carina arm may be part of this
resonance ring.  Key words: {\it Galaxy, spiral pattern,
kinematics and dynamics, Fourier analysis.}
\end{abstract}

\section*{\rm  Introduction}

An analysis of the velocity field of young stars (OB associations
and Cepheids) revealed periodic variations in the radial and
azimuthal components of the residual velocity along the Galactic
radius vector with an amplitude of $f_R=f_\theta=7\pm2$~km
s$^{-1}$ and a scale length of $\lambda=2\pm0.2$~kpc (Mel'nik et
al. 1999, 2001). We determined the kinematic locations of the
Carina, Cygnus, and Perseus arms as the locations of the minima
in the distribution of the radial component of the residual
velocity of OB associations. The identified arm fragments  deviate
from an ideal spiral pattern (smooth trailing spiral arms). The
Carina arm ($R=6.5$~kpc) is displaced from the Cygnus arm
($R=6.8$~kpc) so that together they fall best on the leading
spiral arm. The Perseus arm  unexpectedly terminates, and we see
its extension  in the III quadrant  neither from the kinematics
of young stars nor from the increase in the density of young
objects. An analysis of various spur generation mechanisms showed
that the identified arm fragments are not spurs and are located
near the minimum  of the potential  (Mel'nik 2003). The potential
perturbation itself is probably not  an ideal monochromatic
spiral wave.

Following Kalnajs (1975), let us assume that the spiral
pattern of the Galaxy is a superposition of several spiral waves,
and let us try to identify the various components of its spiral
pattern. Kalnajs (1975) was the first to perform a Fourier
analysis of the spiral pattern of a galaxy using  the distribution
of HII regions in the galaxy M~31. This analysis revealed
a dominat one-armed leading spiral in it.

The spiral pattern of the Galaxy is difficult to analyze, because
we do not have a complete picture of the distribution of young
objects in its disk. On the other hand,  only our
Galaxy makes it possible to identify  arm fragments  kinematically.
Therefore, we have not just a  sample of spiral-arm indicators,
but  a sample of potential-minimum indicators.

\section*{\rm  Fourier Analysis of the Galactic
Spiral Pattern in the Solar Neighborhood }

\subsection*{\it \large The Method and Models}

The complex spiral pattern of any galaxy can be represented as a
superposition of spiral components with different numbers of
arms  $m$, and each $m-$component, in turn, can consist of several
spiral waves with different pitch angles.

An elementary logarithmic spiral wave is defined by the equation
\begin{equation}
R=R_1 e^{\tan i (\theta-\theta_1)},
\end{equation}
where $i$ is the pitch angle, $\theta$ and $\theta_1$ are the
Galactocentric angles, and $R$ and $R_1$ are the Galactocentric
distances.

The amplitudes of the spiral oscillations in the distribution of $N$
objects in the Galactic plane can be determined from the relation
\begin{equation}
A(p,m)=\frac{1}{N} \sum_{j=1}^N e^{-i(m\theta_j+p\ln(R_j/R_0))},
\end{equation}
\begin{center}
where $p=-m/\tan i$
\end{center}
(Kalnajs 1975; Considere and Athanossoula 1982).

Only the interarm distance $\lambda$ and, in certain cases, the
sense of arm winding  can be reliably determined by
analyzing the spiral pattern of the Galaxy within 3~kpc of the
Sun. The number of spiral arms $m$ and their mean pitch angle
$i$ remain indeterminate, although all three parameters are
related by
\begin{equation}
|\tan i|=\frac{\lambda m}{2\pi R_0},
\end{equation}
where $R_0$ is the Galactocentric distance.

The number of spiral arms cannot be determined by analyzing the
distribution of young stars within 3~kpc of the Sun, because all
$m$-components are very similar to one another, and the Fourier
amplitude $A(p,m)$ actually depends only on the  parameter $p$.
Therefore, only the Fourier spectrum of the $m=1$ spiral
component is shown on the plots.

Let us consider  two models of the Galactic spiral arms. In
model~1, the spiral pattern of the Galaxy is represented by 15
real OB associations (Blaha and Humphreys 1989) that are located
near the minimum in the distribution of the radial  residual
velocity $V_R$ and have $V_R<-5$~km s$^{-1}$. Table~1 lists these
associations together with their heliocentric ($r$, $l$, $b$) and
Galactocentric ($R$, $\theta$) coordinates and the residual
velocities $V_R$. Figure~1a shows the locations and residual
velocities of these associations in the Galactic plane and the
circular arcs that correspond to the minima in the distribution
of the radial residual velocity. We excluded from our analysis
the OB associations Sco~OB4 ($l=353^{\circ}$, $r=1.0$~kpc,
$V_R=-15$~km s$^{-1}$) and HD~156154 ($l=351^{\circ}$,
$r=2.1$~kpc, $V_R=-16$ km s$^{-1}$), although their residual
velocities $V_R<-5$~km s$^{-1}$, they deviate greatly from the
observed periodic pattern of the velocity field of young objects.

In model~2, the distribution of young objects is modeled in a
ring. The points representing the OB associations are randomly
scattered over the ring $R_0-2<R<R_0+2$~kpc, except for the 3~kpc
solar neighborhood, where they concentrate toward the three arcs
corresponding to three fragments of the spiral pattern (Fig.~1b).
Table~2 lists the Galactocentric angles of the beginning and end
marks of these arcs $\theta_{\textrm{min}}$ and
$\theta_{\textrm{max}}$ and their Galactocentric distances
$R_{\textrm{arm}}$. The points in the arms are distributed
uniformly, and their radial distribution is limited to the
interval $R=R_{\textrm{arm}}\pm0.$2~kpc. The total number of
points in the arms is equal to the number of points that would
fell within  the selected region of the ring  (within 3~kpc of
the Sun) if it were filled uniformly and  randomly. Of the  2000
objects scattered over the ring, 250 objects, on  average, fall
within the selected region, and  40, 20, and $40\%$ of them fall
within the Carina, Cygnus, and Perseus arms, respectively.

The modeled ring has a width of   $\Delta R=\pm2$~kpc, because
we can ensure the completeness of our sample, or, more specifically,
the identification of all arm  fragments in the solar neighborhood
only in this interval. The tightly wound  spiral pattern
of the Galaxy forces us  to be cautious, since   the density of young objects
changes much faster radially than  azimuthally.

We assumed the distance to the  Galactic center  to be
$R_0=7.1$~kpc (Dambis et al.~1995; Glushkova et al.~1998) and
shorten the heliocentric distances of OB associations given by
Blaha and Humphreys (1989) by 20\% to reduce them to the
so-called short distance scale (Dambis et al.~2001; Sitnik and
Mel'nik~1996).

\section*{\rm  Results}

The heavy line in Fig.~2 indicates the Fourier decomposition of
the distribution of 15 OB associations over the Galactic plane
into logarithmic spiral waves. The  parameter $p$ that
characterizes the interarm distance $\lambda =2\pi R_0/|p|$ and
the sense of arm winding is along the horizontal axis; $p>0$ and
$p<0$ correspond to the trailing and  leading arms, respectively.
The  absolute values of the amplitudes of the one-armed spiral
waves $|A(p,1)|$ are along the vertical axis. Figure~2 clearly
shows three  maxima. The left- and right-most outside maxima at
$p=-28$ and $p=+24$ correspond to a tightly would spiral pattern
with $\lambda=1.8\pm0.4$~kpc. The Fourier spectrum of the tightly
wound spiral arm consists of a superposition of the leading and
trailing components, because we consider only a small region
where the tightly wound spiral arms are indistinguishable from
circular arcs. The circular arc is decomposed into a
superposition of  leading and trailing spiral waves with
approximately equal amplitudes and pitch angles $|i|$. The central
maximum in Figure 2 stems from the fact that we are dealing with
a sample with a highly asymmetric distribution  about the
Galactic center. Indeed, all of the associations in model~1 are
located within 3~kpc of the Sun rather than distributed over the
entire Galactic disk. Such a cluster of objects corresponds to a
spiral wave with an infinite pitch angle ($p=0$). The width of
the central maximum depends on the size of the cluster: the
smaller  the clump, the wider the central maximum.

The dotted line in Fig.~2 indicates the Fourier decomposition
for 15 objects randomly distributed in the 3~kpc solar neighborhood
or, to be more precise, in the region where the circle
$r=3$~kpc and ring $R_0-2<R<R_0+2$~kpc intersect.
We clearly see the maximum at $p=0$ and the amplitude fluctuations  on
both sides of it. The thin line in Fig.~2 indicates the Fourier
decomposition of a random distribution of 15 objects in the above solar
neighborhood  averaged over 100 samples. The Fourier
spectrum of one random distribution agrees well with the
mean spectrum in the central region ($p=0$), where the amplitude
fluctuations are small. The mean spectrum of random distribution
is a symmetric function relative to which we clearly see
the central maximum of the Fourier decomposition of the real sample
to be displaced toward the negative $p$, i. e., toward the leadind waves.
No random fluctuations can explain this displacement.

Let us assume the mean amplitude of the random distribution
of 15 objects to be a background. We are interested in the
difference $\Delta A(p,1)$ between the Fourier amplitude
for the real sample and the background. The heavy line in Fig.~3a
indicates the absolute value of this difference $|\Delta A(p,1)|$.
We clearly see three maxima at $p=-28$,  $p=+24$, and $p=-10$.
The maximum at $p=-10$ indicates the presence of a leading wave with
a large interarm distance of $\lambda=4.5\pm2.0$~kpc in the
morphology of the Galactic spiral pattern.

The rms deviation of the Fourier amplitude of one random
sample from the background is  0.12 and does not depend on
$p$, except for the central region, where it is close to zero
(the thin line in Fig.~3a). The maxima at $p=-28$,  $p=+24$, and
$p=-10$ are not the result of random background fluctuations   at the
$99\%$ confidence level.

Any Fourier decomposition contains high-order harmonics that,  in
our case, correspond to the increasingly  tightly would spiral
patterns at $p=2p_0$, $3p_0$, $4p_0, \ldots$, where $p_0$ is the
fundamental harmonic ($|p_0|=26$). Our  Fourier spectra
also have high-order harmonics, but they are outside our figures.

It is interesting to understand the causes of such deep and sharp
minima in the Fourier decomposition of the real sample (Fig.~2).
They correspond to a wave with an interarm distance
($|p|=13$) that is twice  the main tightly wound spiral pattern ($|p_0|=26$).
Since such a wave would  inevitably arrive  in opposite phases at the
Perseus arm and at the combined Carina--Cygnus arm fragment,
it is absent in the spectrum.

The heavy line in Fig.~3b indicates the Fourier decomposition of the
distribution of young objects in model~2. We clearly see three
maxima at $p=-31$, $p=+26$, and $p=-12$ that  correspond to a tightly
wound spiral pattern with an interarm distance of $\lambda=1.6\pm0.5$~kpc
and a leading wave with $\lambda=3.8\pm2.0$~kpc. No central maximum
emerges in model~2, because any two halves of the Galaxy contain
almost equal number of points. The scales of the vertical axes
of the  models 1 and 2 differ by a factor of 10. However, we are
interested not in the absolute values, but in the form of Fourier
decomposition. The Fourier decomposition of the random distribution
of young objects throughout the ring consists of random amplitude
fluctuations that do not exceed 0.06. The mean amplitude of
these fluctuations is $|A(p,1)|=0.02$ (the thin line in Fig.~3b).
None of the three maxima  at $p=-31$, $p=+26$, and $p=-12$ is the result of
random fluctuations in the distribution of young objects at the  $99\%$
confidence level.

To understand why the leading wave  appears, let us consider the
Fourier decomposition of an ideal spiral pattern in the solar
neighborhood. The model of an ideal spiral pattern is a
modification of model~2 (Fig.~1b). The difference is  that
objects within 3~kpc of the Sun concentrate toward two rather than
three arm fragments and do not terminate anywhere within this
region. The arms are in the shape of circular arcs and are
located at the Galactocentric distances of the Carnina
($R=6.5$~kpc) and Perseus ($R=8.2$~kpc) arms. The model is
symmetric about the $X$-axis. Both arms contain equal numbers of
objects.

The dotted line in Fig.~3b indicates the Fourier decomposition of
an ideal spiral pattern. It exhibits two large maxima at
$p=\pm26$ and two barely visible but unremovable maxima at
$p=\pm9$. The large maxima correspond to a main tightly wound
spiral pattern, while the small maxima correspond to spiral waves
with an interarm distance that is a factor of 3 larger than the
distance between the Perseus and Carina arms. These waves emerge,
because  the region considered includes two arm fragments and the
interarm space between them. Three halves of the main wave,
$3\lambda_0/2$, can be covered by one half of the other wave,
$\lambda_1/2$. This wave describes the behavior of all objects
within 3~kpc of the Sun, without separating them between the
arms, and its maximum must be located approximately halfway
between the arms.

The detected leading wave with $\lambda=4\pm2$~kpc
is of the same nature. It describes the behavior of all
objects within 3~kpc of the Sun. Figure~4 shows the most
probable  location of the leading arm. It has a pitch angle of
$i=6^\circ$ and $i=12^\circ$ for the one-armed ($m_L=1$) and two-armed
($m_L=2$) models of the leading wave, respectively. In both cases,
the leading arm intersects the $X$-axis at a Galactocentric
distance of  $R=7.4\pm0.5$~kpc (for the assumed Galactocentric
distance of the Sun $R_0=7.1$~kpc), i.e., it lies exactly
halfway between the combined Carina--Cygnus  fragment and the
Perseus arm.

Both defects of the spiral pattern --- the displacement of the
Carina and Cygnus arm fragments toward the leading spiral and the
absence of the Perseus arm extension in the  III quadrant ---
give rise to  a leading wave. However, the termination of the
Perseus arm gives a much  larger contribution  to the leading
wave.

Both models yield similar results. However, the following
trend clearly shows up  in model~2. The leading wave
tends to have an interarm distance that is twice $\lambda$ of the
tightly wound spiral pattern. This is because we
actually have two fragments of the spiral pattern: the
combined Carina--Cygnus fragment and the Perseus arm. In order to
weaken the second fragment without weakening the first fragment, the
additional wave must have an interarm distance that is twice that
of the main tightly wound spiral wave. To weaken the Perseus arm in the
III quadrant  without weakening it significantly  in the  II quadrant,
the additional spiral wave must be a leading one.

Thus, the distribution of OB associations within 3~kpc of the Sun
can be represented as a superposition of two spiral components: a
tightly wound spiral wave with an interarm distance of
$\lambda=1.8\pm0.4$~kpc and a leading arm with an interarm
distance of $\lambda=4\pm2$~kpc.

How can the detection of other more distant fragments of the
tightly wound spiral pattern change the parameters of the leading
wave? The leading wave proved to be most sensitive to the
asymmetry of the new possible arm fragments about the $X$-axis.
The amplitude of the leading wave would only increase if, for
example, the new distant fragment detected in the outer Galaxy
were as asymmetric  about the $X$-axis as the Perseus arm (i.e.,
more objects would be located in the  II quadrant than in the III
quadrant, see Fig.~1b). But in the inner Galaxy, a distant
fragment must contain more objects in the  IV quadrant than in
the I quadrant to amplify  the leading wave. In general, the
asymmetry in the distribution of young objects required to
amplify the leading wave agrees with asymmetry in the HI
distribution in the Galaxy (Blitz and Spergel~1991).

 \section*{\rm  How does each Spiral Component
Manifest itself in the Kinematics and Distribution  of Young
Stars?}

While analyzing the locations of spiral arm fragments identified
by the kinematics of young stars, we implicitly assumed that
the additional components of the spiral pattern would not disrupt the
periodic pattern of the velocity field of young stars. The
large interarm distance of the leading component agrees well with
this assumption. Indeed, the velocity perturbation amplitudes
$f_R$ and $f_\theta$ for young stars must  be
proportional to $\lambda^{-1}$ (Lin et al.~1969).
Consequently, even if the other  parameters of the spiral pattern
(the perturbations of the potential and the locations with respect
to the corotation radius) are equal,
the velocity perturbation amplitudes for the   leading wave
must be a factor of 2 lower than those for the tightly
wound wave. In general, the velocity perturbations for young
stars can be explained without invoking the leading wave.

The tightly wound spiral wave was first found by analyzing the
kinematics of young stars (Mel'nik et al.~1999, 2001). In the
Fourier spectrum it is represented by a superposition  of  leading and
trailing spiral waves. However, the kinematics of young objects
strongly suggests that the tightly wound spiral pattern is a
trailing wave.

The sense of winding is determined by the following. The radial
components of the mean residual velocities for the young
stars that concentrate in the Cygnus and Perseus arms are
directed toward the Galactic center, while their azimuthal components
are opposite to the sense of Galactic rotation. Three conclusions
can be drawn from this fact. First, the Cygnus and Perseus arm
fragments represent a trailing spiral pattern. Second, the Cygnus
and Perseus arm fragments are located within the corotation
circle. Third, the coincidence of the minima in the distributions
of  radial and azimuthal  residual velocities
is indicative of the presence of a shock (see Mel'nik (2003) for
detail).

A certain problem arises with the  Carina arm. We observe an enhanced
concentration of young stars in all three arm fragments: the
Perseus, Cygnus, and Carina arms. However, in the Carina arm we
find no significant variations in the azimuthal
residual velocity across the arm, while  the radial  residual velocity
exhibits a well-defined minimum (see Fig.~5 in Mel'nik et al.~2001). The
absence of an azimuthal velocity  gradient across the Carina arm  may be
attributable to distance errors that can effectively blur the variations
exactly  in the azimuthal  velocity (see Mel'nik~2003 for detail).
However, this defect can also have a different cause, which is discussed
in the next section.

The leading wave clearly shows up only in the relative location of
the spiral arm fragments. Indeed, all three fragments
of the Carina, Cygnus, and Perseus arms can be imagined as lying
on a broad leading spiral arm. This distribution can stem from the fact
that the star formation is more intense in the  portions of the
trailing spiral arms, where they crosses the crest of the leading
wave. Shocks seem to be also present only in
these arm portions. This behavior can be explained
using the model of Roberts and Hausman~(1984),
in which  the motion of particle clouds is considered
in a perturbed potential. The enhanced surface
density of the disk in these  portions must cause additional
crowding of the cloud orbits and increase the cloud-cloud collision
frequency, which, in turn, can give rise to shocks.

Thus, the presence of a leading wave in the Galactic disk allows
certain  deviations from an ideal spiral pattern to be explained.
The leading wave emerges mainly because the Perseus arm weakens  in
the   III quadrant. The simplest explanation of  the weakening of
the Perseus arm  consists in a decrease of the disk surface
density in this  region.

\section*{ \rm  Evolution of the Galactic
Spiral Pattern  }

Another peculiarity of the Galactic spiral pattern
is the unusually small interarm distance of the trailing
spiral wave ($\lambda=1.8\pm0.4$~kpc), which is rarely
observed in other galaxies. The tightly wound spiral pattern
of our Galaxy may appear as a ring or  part of it
from a large distance.

The two peculiarities of the Galactic spiral pattern
---  the presence of a shock and a tightly wound spiral pattern
--- force us to abandon the theory of modes, which is attractive in that
it can ensure  a quasi-steady state. The large
energy losses in  the shock require an efficient
spiral wave amplification mechanism from the model to ensure a
quasi-steady state. However, this (swing  amplification) mechanism
gives rise to only open spiral patterns
(Toomre~1981; Athanossoula~1984). The observed
tightly wound spiral pattern is probably  not  a mode and
must evolve along the Galactic radius vector.

Toomre (1969) showed that during several galaxy rotations, the
short trailing spiral wave displaces from the corotation region to
the inner Lindblad resonance. During this displacement, the distance
between the turns of the spiral wave   decreases, while the
density and velocity perturbations of the young stars remain
high for a long time (Toomre~1969, 1977; Lin~1970). The trailing
spiral wave observed in the Galaxy may be located near
the inner Lindblad resonance. The unusually small interarm
distance of the trailing wave can be explained by its evolution:
the decrease in the interarm distance as the wave displaces toward
the inner Lindblad resonance.

The Carina arm may be  part  this resonance ring. First, we
see no tightly wound spiral arms within the region bounded by the
Carina arm. The kinematics of the Sagittarius arm differs
markedly from that of the Perseus, Cygnus, and Carina arms
(see Mel'nik et al.~(2001) for detail) and it cannot be considered
to be the fragment of the same spiral wave. Second, the
ring geometry of the Carina arm could in principle explain
the absence of cross-arm variations in the azimuthal residual velocity.

The superposition of the trailing and leading spiral arms found
in our Galaxy is not unique. The galaxy M~31  exhibits a
similar pattern with the leading and trailing arms having
interarm distances of 6 and 4~kpc, respectively (Considere and
Athanossoula~1982). The main difference is that the interarm distance
of  trailing wave in the Galaxy is a factor of about 2 smaller.
In addition, the leading wave in the Andromeda galaxy is known to be
one-armed. No such information is available for our Galaxy.

\section*{\rm  Conclusions}

We Fourier decomposed  the spiral pattern of the Galaxy
identified by the kinematics of young stars within 3~kpc of the Sun
into spiral harmonics. We showed
that the spiral pattern could be represented as a superposition of
trailing and  leading waves with the interarm distances of
$\lambda=1.8\pm0.4$~kpc and $\lambda=4\pm2$~kpc, respectively.

The presence of a leading wave in the Galactic disk makes it
possible to explain certain deviations from an ideal spiral pattern.
The leading wave clearly shows up only in the large-scale
distribution of young objects and has virtually no effect on
their kinematics. Shocks seem to exist only in the portions of
the trailing spiral pattern where it intersects  the crest of
the leading wave. An enhanced surface density of the disk in these
portions must cause additional crowding of the cloud orbits and
increase the cloud-cloud collision frequency, which, in turn,
can give rise to shocks (Roberts and Hausman~1984).

The small interarm distance of the trailing spiral wave
($\lambda=1.8$~kpc) can be explained by its evolution: the
decrease in the interarm distance as the wave displaces toward the
inner Lindblad resonance. The Carina arm may be  part of this
resonance ring. It may well be that there is a spiral pattern
in the region bounded by the Carina arm,
but this is a different spiral pattern.

The leading wave in our Galaxy can be the source regenerating the
trailing spiral pattern via the swing amplification mechanism
(Goldreich and Lynden-Bell~1965; Julian and Toomre~1966;
Toomre~1981). However, it remains unclear why it emerges.

\section*{\rm  Acknowledgments}

I am grateful to A.V.~Zasov, I.I.~Pasha, A.S.~Rastorgouev,
Yu.N.~Efremov, and A.K.~Dambis for interesting discussions and
useful comments. This work was supported by the Russian
Foundation for Basic Research (projects nos.~02-02-16677 and
03-02-16288), the Council for the Program of Support for Leading
Scientific Schools (projects no.~NSh.389.2003.2), and the
"Astronomy" State Science and Technology Program.

\section*{\rm  References}
\begin{enumerate}

\item E. Athanassoula, Physics Reports {\bf 114}, 319
(1984).

\item C. Blaha and R.M. Humphreys, Astron. J. {\bf 98},
1598 (1989).

\item L. Blitz and D.N. Spergel, Astrophys. J. {\bf 370},
205 (1991).

\item S. Considere and E. Athanassoula, Astron. Astrophys.
{\bf 111}, 28 (1982).

\item  A.K. Dambis, A.M. Mel'nik, and A.S. Rastorguev,
 Astron. Letters, {\bf 21}, 291 (1995).

\item  A.K. Dambis, A.M. Mel'nik, and  A.S. Rastorguev,
  Astron. Letters, {\bf 27}, 58 (2001).

\item E.V. Glushkova, A.K. Dambis, A.M. Mel'nik, and A.S.
Rastorguev, Astron. Astrophys. {\bf 329}, 514 (1998).

\item P.Goldreich and D. Lynden-Bell, MNRAS {\bf 130}, 125
(1965).

\item W.H. Julian and A. Toomre, Astrophys. J. {\bf 146},
810 (1966).

\item A.J. Kalnajs, {\it La Dynamique des Galaxies Spirales,
Coll. Int. CNRS No 241} (Ed. L. Weliachew, Paris: Editions du
Centre National de la Recherche Scientifique, 1975), p. 103.

\item C.C. Lin,  {\it The Spiral Structure of Our Galaxy, IAU
Symp. {\bf 38}} (Ed. W. Becker and G. Contopoulos, Dordrecht:
Reidel, 1970), p.377.

\item C.C. Lin, C. Yuan, and F.H. Shu, Astrophys. J. {\bf
155}, 721 (1969).

\item  A.M. Melnik,  Astron. Letters, {\bf 29}, 304 (2003).

\item  A.M. Mel'nik, A.K. Dambis,  and A.S. Rastorguev,
 Astron. Letters, {\bf 25}, 518 (1999).

\item  A.M. Mel'nik, A.K. Dambis, and  A.S. Rastorguev,
 Astron. Letters, {\bf 27}, 521 (2001).

\item W.W. Roberts, and M.A. Hausman, Astrophys. J. {\bf
277}, 744 (1984).

\item T.G. Sitnik  and A.M. Mel'nik , Astron. Letters, {\bf
22}, 422 (1996).

\item A. Toomre, Astrophys. J. {\bf 158}, 899 (1969).

\item A. Toomre, Ann. Rev. Astron. Astrophys. {\bf 15}, 437
(1977).

\item A. Toomre, {\it The Structure and Evolution of Normal
Galaxies }(Ed.  S.M. Fall and D. Lynden-Bell, Cambridge: Cambridge
University Press, 1981) p. 111.
\end{enumerate}

\newpage

\makeatletter
\renewcommand{\@oddhead}{}\begin{center}
\makeatother

Table 1. List of OB associations used in model~1\\[10pt]

\begin{tabular}{l|c|c|c|c|c|c}
\hline
OB association& $l$ & $b$ & $r$ & $V_R$ & $R$ & $\theta$\\
           &deg. & deg.& kpc & km s$^{-1}$& kpc & deg.\\[5pt]
\hline
Cyg OB3&      72.8&  2.0&  1.8&   -14.5&    6.79&  14.93\\
Cyg OB9&      77.8&  1.8&  1.0&    -5.8&    6.96&   7.74\\
Cep OB1&     104.2& -1.0&  2.8&    -8.7&    8.23&  19.08\\
Cas OB2&     112.0&  0.0&  2.1&   -19.0&    8.13&  13.89\\
Cas OB5&     116.1& -0.5&  2.0&   -12.7&    8.18&  12.73\\
Cas OB7&     123.0&  1.2&  2.0&   -10.3&    8.36&  11.61\\
Cas OB1&     124.7& -1.7&  2.0&    -6.8&    8.41&  11.31\\
Per OB1&     134.7& -3.2&  1.8&    -7.8&    8.49&   8.81\\
Cas OB6&     135.0&  0.8&  1.8&   -11.2&    8.43&   8.44\\
Coll 228&    287.6& -1.0&  2.0&    -6.6&    6.77& -16.42\\
Car OB2&     290.4&  0.1&  1.8&    -5.6&    6.69& -14.20\\
Cru OB1&     294.9& -1.1&  2.0&   -13.5&    6.51& -16.23\\
NGC 3766&    294.1& -0.0&  1.5&    -7.3&    6.62& -12.15\\
Cen OB1&     304.2&  1.4&  1.9&   -16.8&    6.23& -14.77\\
Hogg 16&     307.5&  1.4&  1.5&    -7.4&    6.32& -10.52\\
 \hline
\end{tabular}

\vspace{3cm} Table 2. Parameters of the spiral pattern used in
model~2\\[10pt]
\begin{tabular}{l|c|c|c}
\hline Arm fragments & $\theta_{min}$ & $\theta_{max}$ &
$R_{arm}$\\
&  deg.& deg.&  kpc \\[5pt]
\hline
       The Carina arm &            -25&     - 5&     6.5\\
       The Cygnus arm &           + 5&     +25&     6.8\\
       The Perseus arm &           + 5&     +21&     8.2\\
\hline
\end{tabular}
\end{center}

\newpage
\begin{figure}[t]
\includegraphics{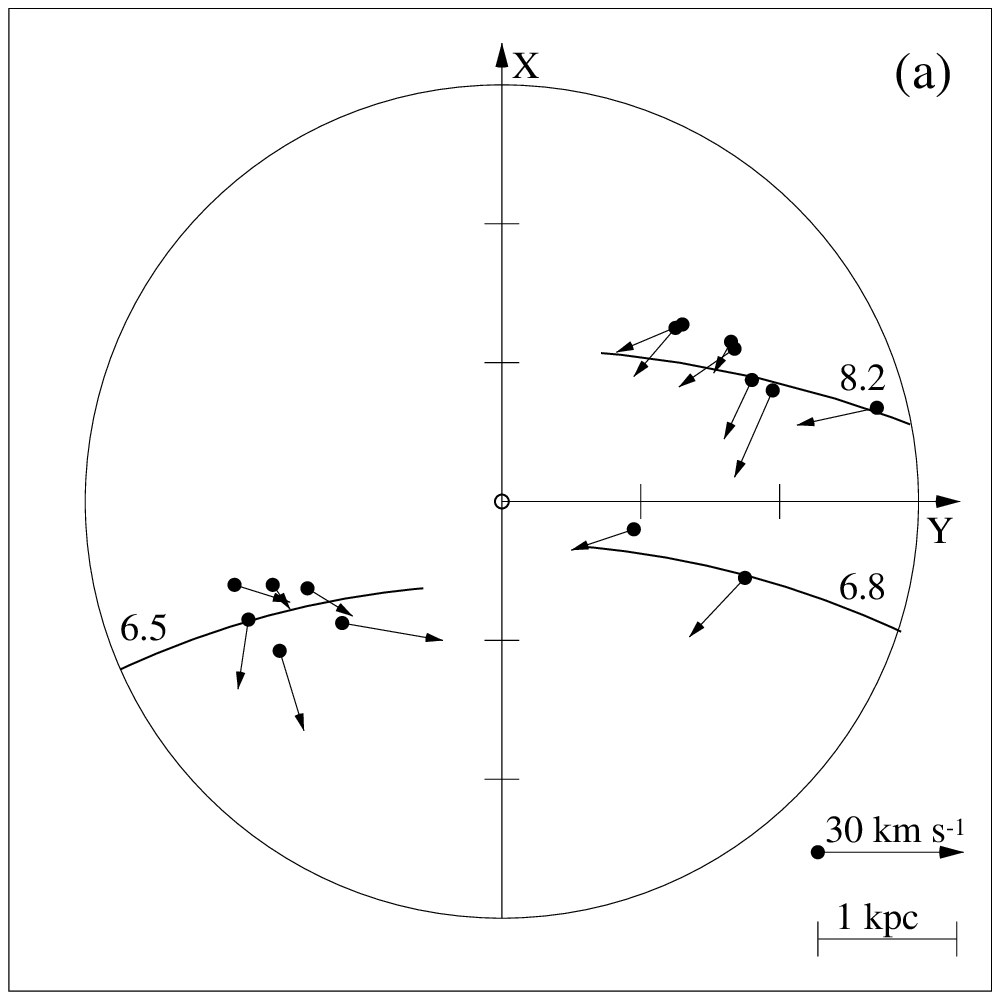} \includegraphics{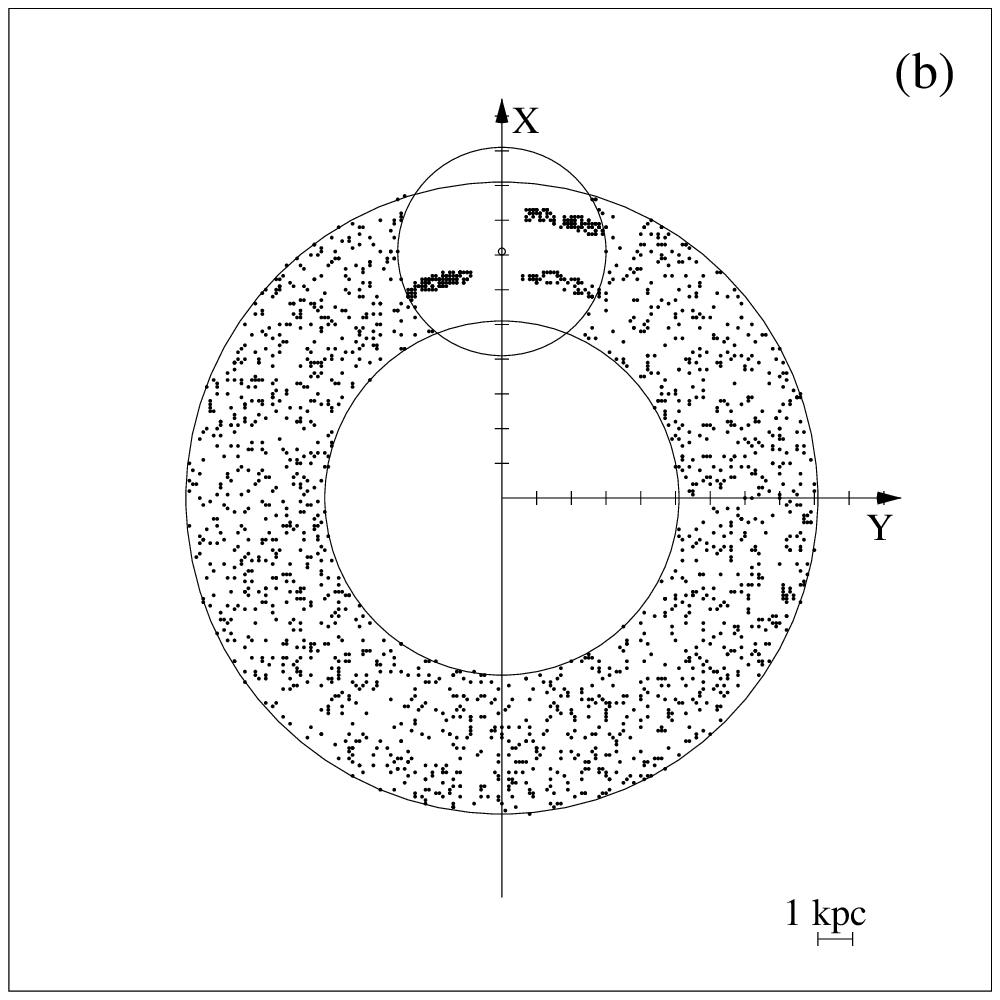} \vspace{14.0cm} \caption{Distribution of
young objects in the Galaxy. (a) In model~1,  the spiral pattern
is represented by 15 real kinematically identified
OB associations. The circular arcs correspond to the kinematic
locations of the spiral arms. (b) In model~2, the  young objects are
randomly distributed in the ring $R_0-2<R<R_0+2$~kpc, except for
the 3~kpc solar neighborhood where they concentrate toward the
identified fragments of the spiral pattern. \hfill}
\end{figure}

\newpage
\begin{figure}[t]
\includegraphics{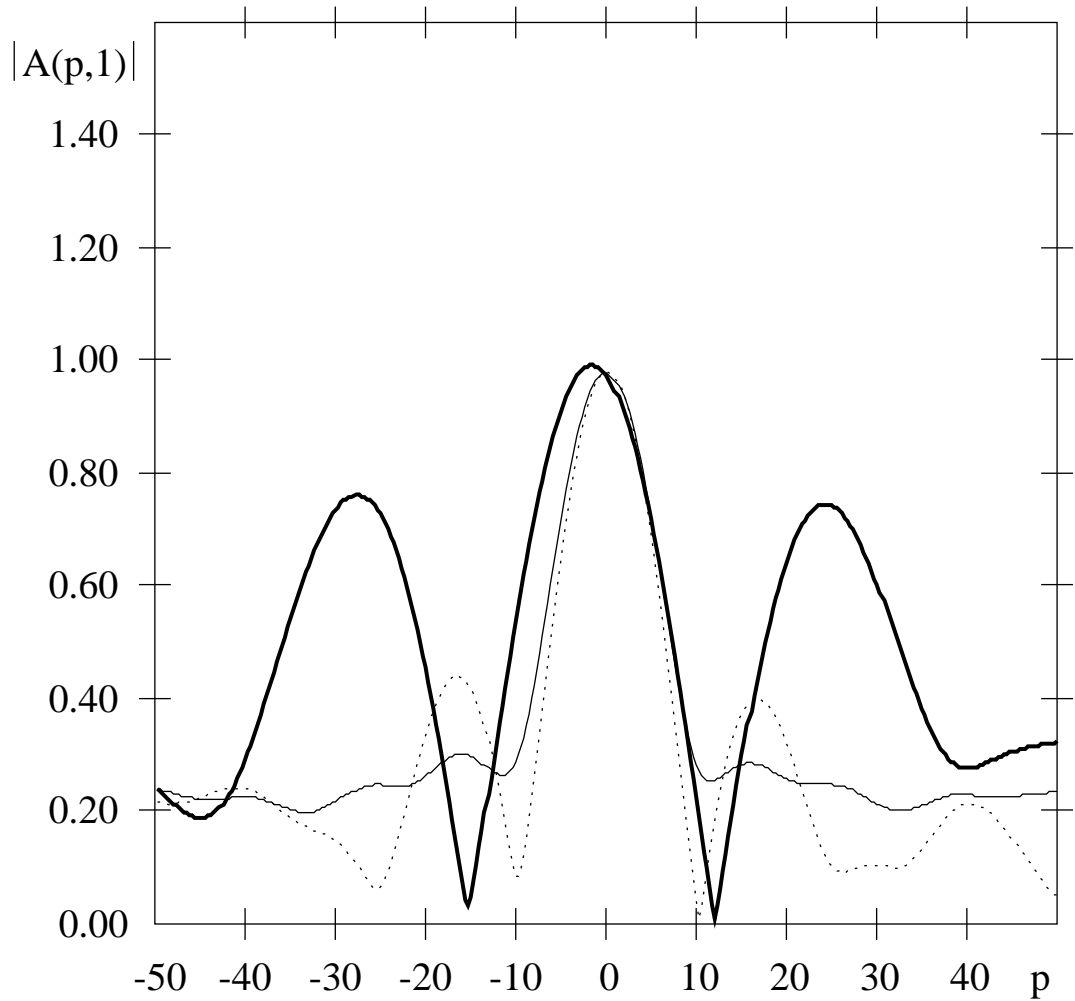}  \vspace{12.0cm} \caption{Fourier decomposition of
the distribution of 15 real OB associations into spiral harmonics
(heavy line); the Fourier spectrum of one random distribution of 15
objects (dotted line), and the Fourier decomposition of the random
distribution averaged over 100 samples (thin solid line)}
\end{figure}

\newpage
\begin{figure}[t]
\includegraphics{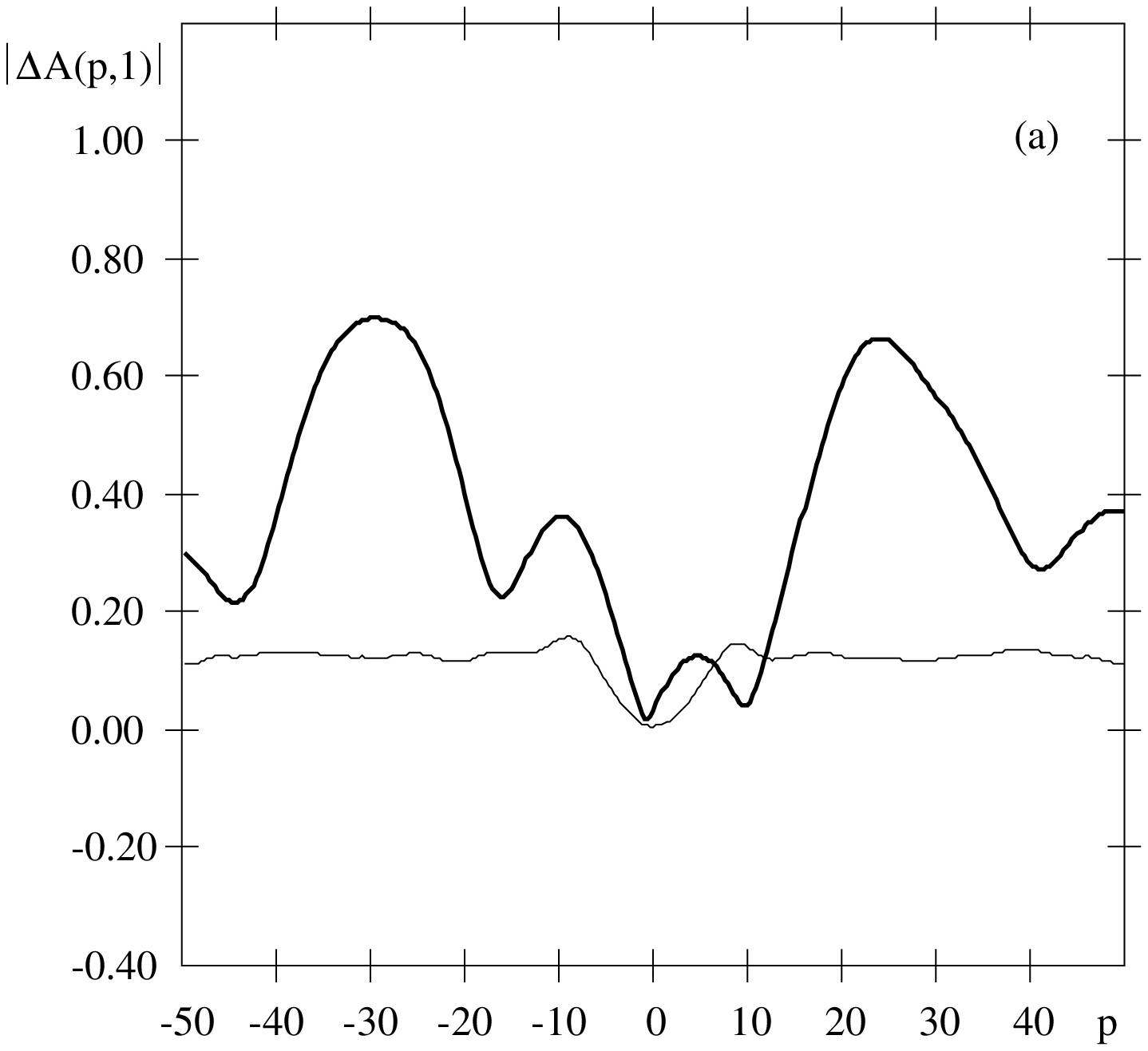} \includegraphics{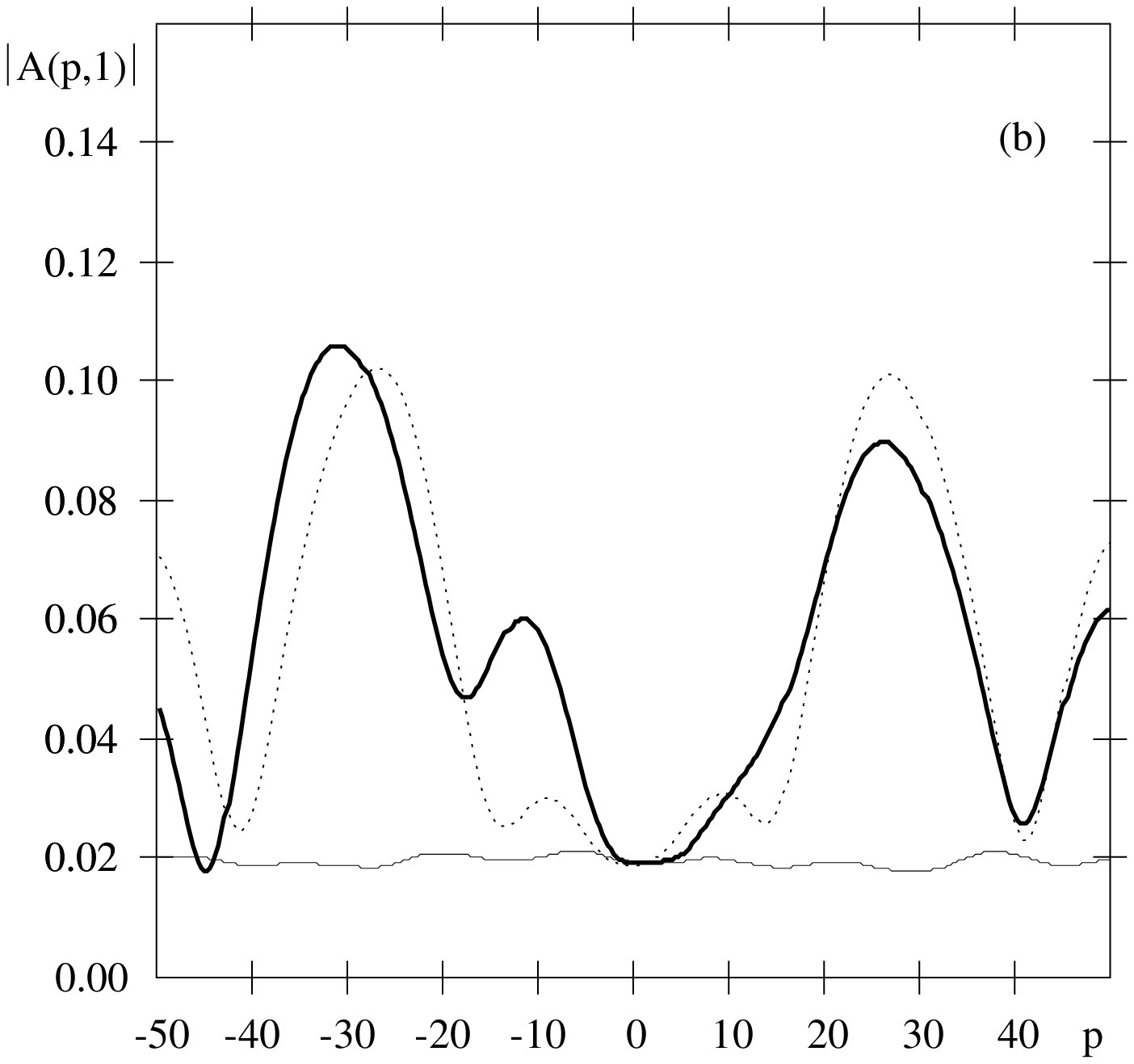} \vspace{14.0cm} \caption{(a)  Absolute value
of the difference between the Fourier amplitude calculated for
the real distribution and the mean Fourier amplitude for a random
distribution of 15 objects (heavy line); the rms deviation of the
amplitude of a random sample from the background level (thin
line). (b) The Fourier decomposition of the distribution of young
objects in model~2 (heavy line); the  Fourier decomposition for
the fragments of an ideal spiral pattern (dotted line);  the mean
amplitude of the random distribution of young objects in the ring
(thin solid line). \hfill}
\end{figure}

\newpage
\begin{figure}[t]
\includegraphics{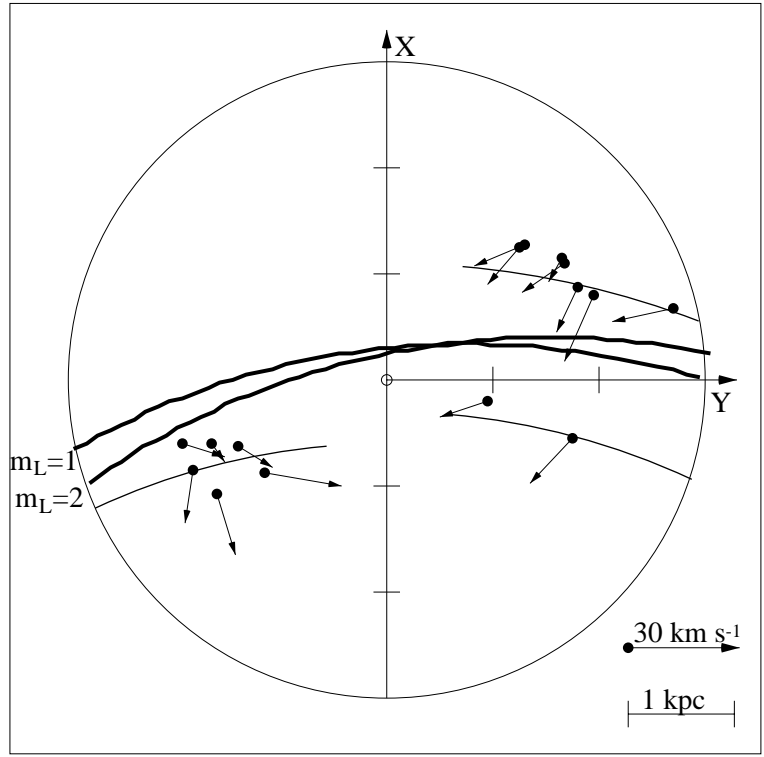}  \vspace{12.0cm} \caption{Most probable locations of
the leading wave with  $\lambda=4.5$~kpc
and a pitch angle of $i=6^\circ$ or $i=12^\circ$ for the one-armed
($m_L=1$) and two-armed ($m_L=2$) models of the leading wave.  In
both cases the leading wave in the solar vicinity  is lying in
nearly between the combined Carina-Cygnus and Perseus arm
fragments.\hfill}
\end{figure}

\end {document}